\documentclass[12pt]{article}
\usepackage[left=2cm,top=2cm,right=2cm,bottom=2cm]{geometry}
\usepackage{authblk}
\usepackage{amsfonts,graphicx,epsfig,epstopdf,hyperref}
\usepackage{amssymb,amsmath,enumerate,float,amsthm}
\usepackage{appendix}
\usepackage{url}
\graphicspath{{./figures/}}
\usepackage{color}
\newcommand{\ptl}{\partial}
\newcommand{\lx}{s}
\newcommand{\ly}{q}
\newcommand{\lk}{{\tilde k}}

\newcommand{\rmU}{{\rm U}}
\newcommand{\rmV}{{\rm V}}
\newcommand{\rmA}{{\rm K}}
\newcommand{\rmF}{{\rm F}}
\newcommand{\rmX}{{\rm X}}

\newtheorem{rem}{Remark}
\newcommand{\bnu}{{\boldsymbol{\nu}}}
\newcommand{\bmu}{{\boldsymbol{\mu}}}

\title{Embedding formulae for diffraction problems on square lattices}
\author{A. I. Korolkov, A. V. Kisil}

\begin{document}

\maketitle
\begin{abstract}
We develop embedding formulae for all possible diffraction problems with Dirichlet scatterers on square lattices using the Wiener–Hopf perspective. The embedding formula expresses solutions for arbitrary plane-wave incidence in terms of a finite set of auxiliary problems, eliminating the need to re-solve boundary value problems for each incidence angle. First we derive explicit embedding formulae for canonical geometries including the half-plane, finite strip, and right-angled wedge. We then generalize the method through an operator-based approach, obtaining embedding formula for arbitrary configurations of obstacles on lattices. This general embedding formula is a key difference from the continuous setting where this is currently not possible. To validate the theory, we perform numerical experiments, confirming agreement with the results derived using the embedding formula. The results highlight the efficiency and generality of the Wiener–Hopf approach in discrete diffraction theory, with potential applications in inverse problems and other areas of physics and mathematics.
\end{abstract}

\section{Introduction}

Recently, waves on lattices have been actively investigated. These problems have applications in many areas: in photonics \cite{Zheludev2010}, where obstacles are interpreted as defects in crystals, in numerical analysis \cite{Finite_difference_book}, where lattice waves approximate the continuos ones as the lattice spacing goes to $0$, fracture mechanics \cite{Mass_spring_review}, where lattices are interpreted as spring-mass systems. There is extensive current research into more advanced lattices and metamaterials~\cite{Nieves_21, MovchanBook,Craster_16}. Additionally, many analytic results were derived using Wiener--Hopf method \cite{Sharma2015,Kisil2024,Medvedeva24,Nieves_2d_24} and other methods of analytic diffraction theory \cite{Shanin2020,Shanin2022}.
Hence, the results of the current study have a wide range of potential applications and generalizations.

The embedding formula is a mathematical relation that expresses the solution of a given  problem in terms of the solution of a set of reference problems. In diffraction theory, it is an algebraic relation that allows to recover the solution for a whole family of plane wave incidence problems from a small set of auxiliary solutions, typically edge Green's functions \cite{Craster2003} or other plane wave solutions \cite{Biggs2006}. In this work, we focus on far‑field patterns (directivities). Once the auxiliary directivities are available, the directivity for any incidence angle can be constructed as a linear combination with angle‑dependent coefficients, eliminating the need to re‑solve the full boundary‑value problem. An exhaustive review of the topic can be found in \cite{Korolkov2024}. 
\begin{figure}
    \centering
    \includegraphics[width=0.6\linewidth]{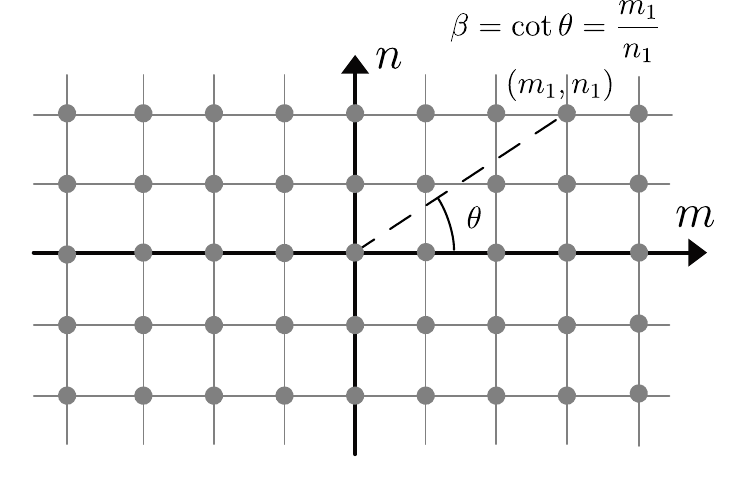}
    \caption{Geometry of the lattice}
    \label{fig:grid}
\end{figure}

\begin{figure}
    \centering
    \includegraphics[width=0.49\linewidth]{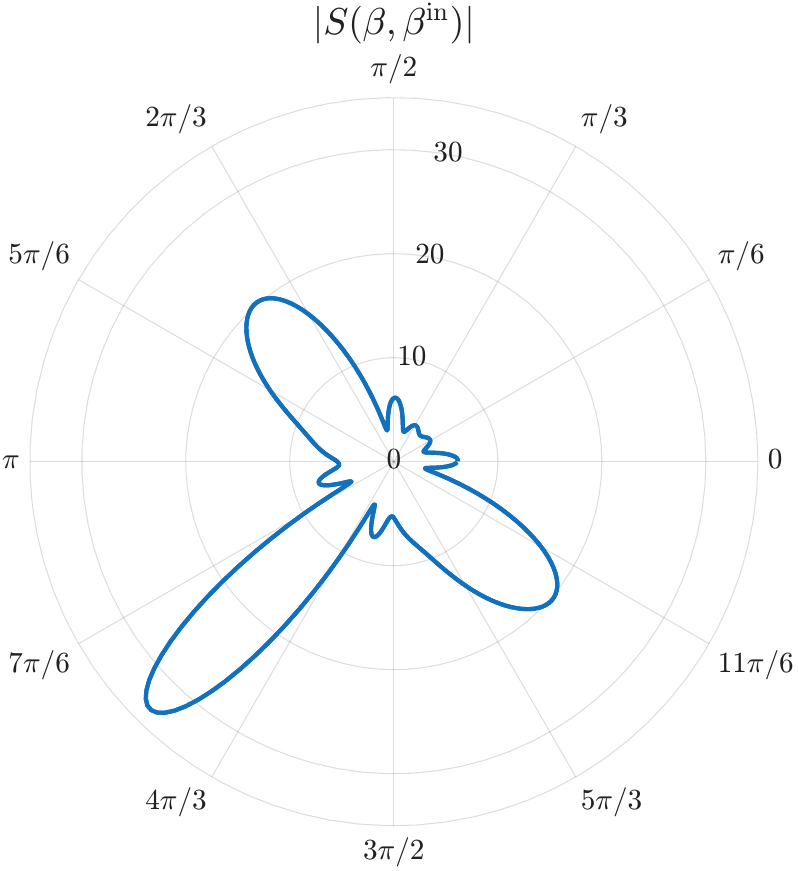}
    \includegraphics[width=0.49\linewidth]{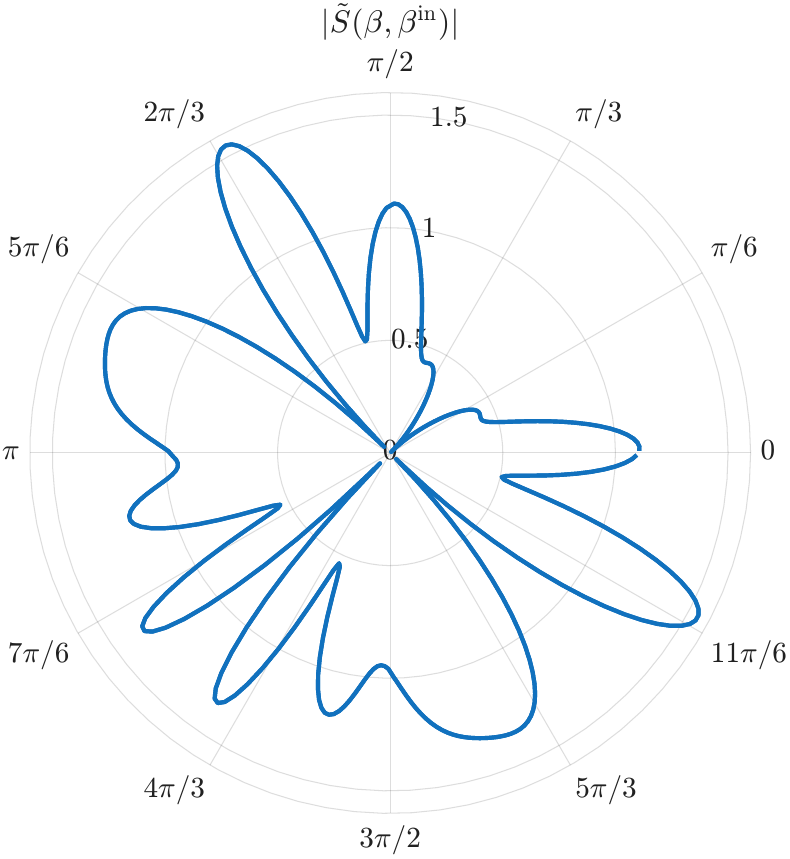}
    \caption{Absolute value of the directivity (left) and the modified directivity (right) for the problem of diffraction by a square with side length $21$ nodes, with $\beta^{\rm in} =1$ and $\tilde k = 0.6+ 0.01i$}
    \label{fig:dir_mod_dir}
\end{figure}

The key concept in this paper is to define the modified directivity for the lattice scattering problems.  The directivity of the scattered field produced by a plane wave, \(S(\beta,\beta^{\rm in})\), can be defined for lattice problems much in the same way as directivity in the continuous case, see Section~\ref{subsec:directivity}. The main difference is that instead of parametrising by an angle \(\theta\) on a lattice it is convenient to use  \(\beta=\cot\theta=m_1/n_1\), see~Figure~\ref{fig:grid}. The modified directivity is defined as 
\begin{equation}
\label{eq:Mod_Dir_wedge}
\tilde S(\beta,\beta^{\rm in}) = (s_\beta+ s_\beta^{-1} - s^{\rm in}-(s^{\rm in})^{-1})S(\beta,\beta^{\rm in}).
\end{equation}\
where \(s_\beta\) is a function of \(\beta\) and \(s^{\rm in}\) is a function of \(\beta^{\rm in}\), see Remark~\ref{rem:label2}. In Figure~\ref{fig:dir_mod_dir} we present absolute value of the directivity and the modified directivity for the problem of diffraction by a square. Further details are provided in Section~\ref{sec:numerics}.

\begin{figure}
    \centering
    \includegraphics[width=0.9\linewidth]{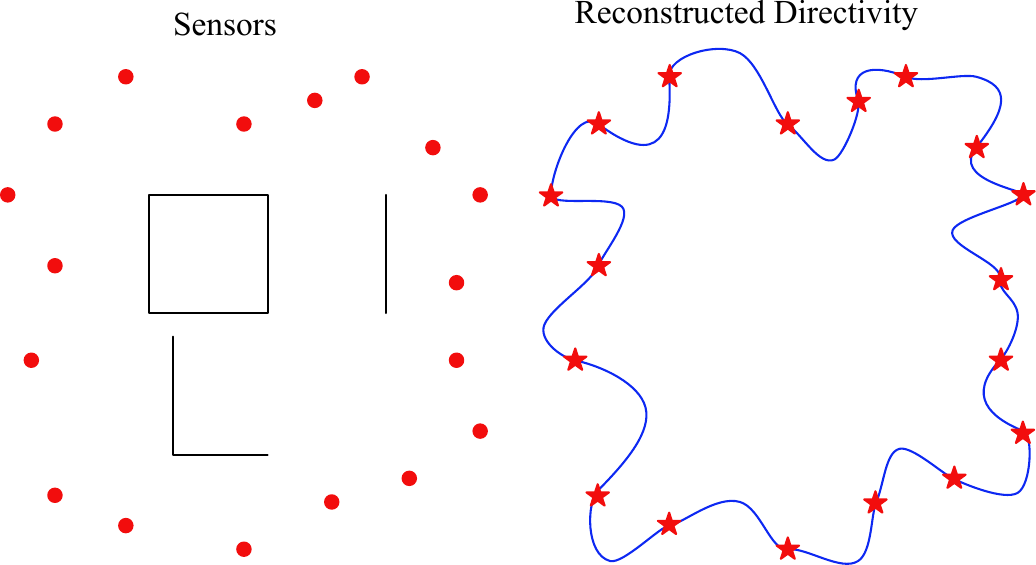}
    \caption{The figure on the left shows the possible set up where the directivity is measured at N points. The figure on the right sketches the whole directivity reconstructed only by the information given by the stars, which are measured at the sensors}
    \label{fig:Dir_reconstruction}
\end{figure}

The advantage of defining the modified directivity is the main result of this paper, \textit{the embedding formula for directivities}
\begin{equation}
\label{eq:weak_embeding_int}
\tilde S(\beta,\beta^{\rm in}) = \sum_{l=1}^NA_l\tilde S(\beta,\beta_l^{\rm in}),
\end{equation}
where \(A_l\) are coefficients and \(N\) is twice the number of all corners in the scatterers. This formula is derived in Section~\ref{sec:embed_gen}. This key difference to the other known embedding formulae for continuous diffraction problems is that for square lattices there is a way to capture all possible Dirichlet obstacles. This insight gives a number of advantages.
\begin{enumerate}
    \item Only \(N\) directivities need to be calculated and then for all other incident angles the directivity can be recovered saving time and/or memory. In fact, \(A_l\) could also we recovered using reciprocity leading to the \textit{strong embedding formula}. For examples, see Section~\ref{sec:numerics}
    \item If a modified directivity produced by a plane wave is known at \(N\) points only then it can be recovered everywhere using \eqref{eq:weak_embeding_int}, if \(\tilde S(\beta,\beta_i^{\rm in})\) for  \(i=1 \dots N\) is known, see Figure~\ref{fig:Dir_reconstruction}.
    \item If \(N\) is not known and only \(\tilde S(\beta_m,\beta_l^{\rm in})\) is known for \(m,l=1,2 \dots M\) for \(M\) large enough, then \(N\) can be recovered, giving information about the scatterers involved in producing this directivity. See Figure~\ref{fig:ranks} in Section~\ref{sec:numerics}. 
\end{enumerate}

In order to derive these embedding formulae we use an approach based on the Wiener--Hopf formulation of the problem, proposed in \cite{Korolkov2024}. The advantage of this approach is that, once the Wiener–Hopf equations are obtained, the derivations reduce to simple algebraic manipulations. Therefore, it allows an effortless derivation of embedding formulae in any context. Additionally, in the recent study \cite{Korolkov2025}, the authors developed a method of analogies that allows a straightforward derivation of the Wiener-Hopf equation for the lattice problem, starting from the continuous counterparts. Hence, the derivations of the current study are simplified by the machinery developed in these works and demonstrate the power of the Wiener--Hopf approach to embedding. Furthermore, in Section~\ref{sec:embed_gen} we use the results derived for canonical problems to introduce an operator approach to embedding on lattices for arbitrary scatterers. 

The paper is organised as follows. Section~\ref{sec:preliminaries} collects some prerequisites which are different for lattice problems compared to continuous problems and  the Wiener--Hopf approach to embedding is recapped in Section~\ref{sec:WH_embedding}. In Section~\ref{sec:canonical}, we formulate a discrete diffraction problem by the half-plane, the strip and the wedge, with an outline of the derivation of the embedding formula. In Section~\ref{sec:embed_gen} we explore the operator approach to embedding and derive the embedding formula for an arbitrary obstacle on a square lattice. In Section~\ref{sec:numerics}, we provide numerical demonstrations. The solutions of several diffraction problems are built using the method of boundary algebraic equations~\cite{Martinsson2009,PobletPuig2018}, and then it is checked directly that the corresponding embedding formulae are satisfied. Concluding remarks are provided in Section~\ref{sec:conclusion}. 

\section{Preliminaries}
\label{sec:preliminaries}

\subsection{Governing equation}
Let us consider a uniform square lattice numbered with integers \((m, n)\), see~Figure~\ref{fig:grid}.
Define a homogeneous discrete Helmholtz equation on a lattice in the usual way
\begin{equation}
\label{eq:Helmholtz}
\Delta_{(m,n)}[u] + \lk^2 u(m,n) = 0,
\end{equation}
with \(\lk\) being the wavenumber which is taken to have a small positive imaginary part, introducing attenuation in the lattice. 
 The operator $\Delta_{(m,n)}[\cdot]$ is the 2D discrete Laplace operator, which is a 5-point finite difference approximation of the continuos Laplace operator:
\[
\Delta_{(m,n)}[u] = u(m,n+1) + u(m,n-1) + u(m+1,n) + u(m-1,n)-4u(m,n).
\]
\subsection{Plane wave}
\label{subsec:plane wave}
We can define a plane wave on a lattice as
\begin{equation}
\label{eq:inc_wave1}
u^{\rm in}(m,n) = \lx^{m} \ly^{n},
\end{equation} 
substituting this into \eqref{eq:Helmholtz} implies that $\lx$ and $\ly$ are solutions to the dispersion equation
\begin{equation}
\label{eq:lattice_Disp_eq}
D_d(\lx,\ly) = \lx+ \lx^{-1} + \ly+\ly^{-1} + \lk^2-4 = 0,
\end{equation}
note that if \(\lx\) is a solution then \(\lx^{-1}\) is also a solution and that there are no solutions with \(|\lx|=1\) since \(\lk^2\) has a non zero imaginary part. Define
\begin{equation}
\label{eq:real_waves_cond}    
\beta = \frac{s - s^{-1}}{q - q^{-1}}.
\end{equation}
This expression appears in the steepest descent analysis in determining the asymptotic estimates  of the Green's function, see  Appendix~\ref{app:app_Green}.
In what follows we will only be interested in solutions  \(\lx\) and \(\ly\) of \eqref{eq:lattice_Disp_eq} so that \(\beta\) is real, this ensures that 
\[|\lx|, |\ly| \to 1 \quad \text{as} \quad \operatorname{Im}(\lk) \to 0\]
and hence the waves are non-decaying  for \(\lk\) real \cite{Shanin2020}. 
 In fact $\beta$ is a real rational number that is linked to the angle of incidence of the plane wave: $\cot\theta =\beta$. 

\begin{rem}
In order to see the relation with the incident angle note that \[\lx^{m} \ly^{n}=\exp^{(i m h\xi \cos(\theta)+ i n h\xi \sin(\theta))},\]
where $h$ is the length of lattice edges. It implies that for $m_1$ and $n_1$ laying on the wavefront
\begin{equation}
\label{eq:plane} 
\lx^{n_1}=\ly^{m_1}.
\end{equation}
Indices $m_1$ and $n_1$ can be linked to the angle of incidence as follows
\[\sin(\pi/2+ \theta)=\frac{n_1}{\sqrt{n_1^2+m_1^2}}, \quad \cos(\pi/2+\theta)=\frac{m_1}{\sqrt{n_1^2+m_1^2}}.\]
Using that 
\[\frac{dq}{ds} = -\frac{q(s-s^{-1})}{s(q-q^{-1})},\]
 differentiating \eqref{eq:plane} with respect to \(s\) leads to (\ref{eq:real_waves_cond}).
\end{rem}
\begin{rem}
\label{rem:label2}
For a given $\beta$, equation (\ref{eq:real_waves_cond}) can be reduced to a fourth-order polynomial equation in $s$, with two solutions corresponding to non-real waves and two corresponding to incoming and outgoing real waves. We denote the solution corresponding to an outgoing wave as $s_\beta$.
\end{rem}
\subsection{ Far Field Directivity}
\label{subsec:directivity}

Define the directivity of the scattered field (which is the total field minus the incident plane wave with parameter \(\beta^{\rm in}\)) as follows:
\begin{equation}
\label{eq:Far_field}
u^{\rm sc}(m,n) = g(m,n)S(\beta,\beta^{\rm in}) + O(N^{-3/2}),\quad N = \sqrt{m^2 + n^2}, \quad \beta=\frac{m}{n},    
\end{equation}
where $g(m,n)$ (derived in Appendix~\ref{app:app_Green}) is the saddle point asymptotic of the free space Green's function.

The main application of the embedding formula in this paper concerns the far field, since the entire field can be reconstructed from the far field via Green’s reconstruction formula (\ref{eq:rec_formula}) and the equation that relates the directivity and the field spectrum on the surface of the scatterer (see~Appendix~\ref{app:Directivity}), in the same manner as in continuous diffraction theory. 

\subsection{Reciprocity relation on a lattice}
It is well known that the reciprocity relation holds for continuous Helmholtz equation and this is also true for the discrete Helmholtz equation. We will need the reciprocity for  directivities, 
\begin{equation}
\label{eq:recipr_dir}
S(\beta,\beta^{\rm in}) = S(\beta^{\rm in},\beta), 
\end{equation}
see Appendix~\ref{app:reciprocity} for the derivation using Green's identity. The reciprocity is key for finding the coefficients in \eqref{eq:weak_embeding_int} and is used to obtain a linear system of equations to solve for the coefficients \eqref{eq:srong_der_embeding}. Note that in Section~\ref{sec:canonical} for the canonical problems symmetry considerations in Fourier space were used instead of reciprocity, and this is equivalent.

\subsection{Wiener--Hopf approach to embedding}
\label{sec:WH_embedding}
In the following section canonical diffraction problems will be reduced to the Wiener--Hopf equation~\cite{WHreview21}, which will give a routine way of obtaining the embedding formula~\cite{Korolkov2025}. 
This method will be summarised here.  Let the problem under consideration be formulated as the following inhomogeneous Wiener--Hopf equation
\begin{equation}
\label{eq: matrix_WH}
\rmU^-(t,z_i)  = {\rmA}(t) {\rmU}^+(t,z_i) + \rmF(t,z_i),\quad t \in     \mathbb{L},
\end{equation}
where $\mathbb{L}$ is a closed curve on the compactified complex plane, and ${\rmU}^+(z,z_i)$ and ${\rmU}^-(z,z_i)$ are vector functions of size $N$ analytic inside and outside the domain restricted by $\mathbb{L}$, ${\rmA}(z)$ is known matrix function of size $N\times N$ that referred to as the kernel of the Wiener--Hopf equation, ${\rmF}(t,z_i)$ is a vector  of forcing terms of size $N$, and $z_i$ is a complex parameter. Specifically, we are interested in the particular form of forcing
\begin{equation}
  \label{eq:pole_forcing}
{\rmF}(z,z_i) = \frac{{\rm r}}{z-z_i},\quad {\rm Im}[z_i]<0,
\end{equation}
where ${\rm r}$ is some constant vector. In the context of the discrete diffraction theory, (\ref{eq:pole_forcing}) corresponds to forcing by a plane wave incidence, $z_i$ is associated with the angle of incidence, and  $\mathbb{L}$ is the unit circle.

The equation \eqref{eq: matrix_WH} can be solved formally using the method of normal solutions \cite{gakhov1952riemann}. Consider a homogeneous Wiener--Hopf problem, i.e (\ref{eq: matrix_WH}) with $\rmF={\rm 0}$:
\begin{equation}
\label{eq: Wh_homo}
{\rmV}^-_{(j)}(t)={\rmA}(t){\rmV}^+_{(j)}(t),\quad t \in \mathbb{R}, \quad j = 1,\ldots,N.
\end{equation}
Note that solutions $\rmV^\pm_{(j)}$ do not depend on parameter $z_i$.  A system of $N$ solutions to the above  $\rmV^{\pm}_{(n)}(z)$, \(n=1, \dots, N\) is called \textit{normal}  if matrix
\begin{equation*}
{\rmX}^{\pm}(z) = ({\rmV}^{\pm}_{(1)}(z),\ldots,{\rmV}^{\pm}_{(N)}(z))
\end{equation*}
does not have zeroes of its determinant on the whole complex plane (except possibly at infinity). Matrix ${\rmX}$ is referred to as normal. The normal matrix ${\rmX^{\pm}}$ provides a multiplicative factorisation of ${\rmA}(t)$ since
 \begin{equation}
 \label{eq: WH_fact}
 {\rmA}(t) = {\rmX}^-(t)({\rmX}^+(t))^{-1}.
 \end{equation}
 Then, by rearranging the equation \eqref{eq: matrix_WH} into two groups of terms analytic inside and outside of $\mathbb{L}$, respectively, and applying Liouville's theorem, obtain
 \begin{equation}
 \label{eq: WH_spliting}
{\rm U}^{\pm}(z,z_i) = {\rm X}^{\pm}(z)\left({\rm P}(z)\mp [\left({\rm X}^-(z)\right)^{-1}{\rm F}(z,z_i)]_{\pm}\right), 
 \end{equation}
   where
   \[\left({\rm X}^-(z)\right)^{-1}{\rm F}(z,z_i)=\left[\left({\rm X}^-(z)\right)^{-1}{\rm F}(z,z_i)\right]_-+\left[\left({\rm X}^-(z)\right)^{-1}{\rm F}(z,z_i)\right]_+,\]
 is the additive Wiener--Hopf splitting which can be computed using a Cauchy type integral,  and ${\rm P}(z)$ is a polynomial vector-function with arbitrary coefficients,  the order of which is determined by the growth rate of $\rmU^\pm$.  Below, we suppose that  ${\rm P}(z)=0$, which holds for the problems considered in this study.
 
The case of pole forcing (\ref{eq:pole_forcing}) leads to considerable simplifications:
\begin{equation}
\label{eq:Can_solution_pole2}
{\rmU}^{+}(z,z_i) = -\frac{{\rm X}^{+}(z)({\rmX}^-(z_i))^{-1}{\rm r}}{z-z_i}. 
\end{equation}
This equation is written as a combination of functions not dependent on \(z_i\) and constants which depend on \(z_i\). In other words, it is an embedding formula for the Wiener--Hopf equation which does not involve any Cauchy integration. This procedure can be generalised to any forcing imposed by a rational function, and even for some irrational forcing terms (see section~4.3 in \cite{Korolkov2024}). To simplify things even further, we will use plane wave solutions to construct the normal solutions, which leads to the plane wave embedding formula \cite{Biggs2006}. 

Indeed, consider a set of $N$ solutions ${\rmU^\pm(z,z_l)}$ of (\ref{eq: matrix_WH}) with pole forcing (\ref{eq:pole_forcing}) for different values of parameter $z_l$ and index them as  $(l= 1, \dots,N)$.  The set is not normal, since solutions have poles that correspond to the forcing. Let us modify this  set as follows: 
\[
\hat \rmU^\pm(z,z_l) = (z-z_l)\rmU^\pm(z,z_l).
\]
The matrix 
\[
\rmX^\pm_p(z) = \left(\hat \rmU^\pm(z,z_1),\ldots,\hat \rmU^\pm(z,z_N)\right)
\]
composed of modified solutions satisfies the following equation: 
\[
\rmX^-_p(z)  - {\rm R}= \rmA\rmX^+_p(z), ,\quad {\rm R} = ({\rm r},\ldots,{\rm r})
\]
Assuming that ${\rm det}(\rmX^+_p)$ and  ${\rm det}(\rmX^-_p(z)  - {\rm R})$ are non-zero we can take 
$
\rmX^-_p(z)  - {\rm R}
$
as the matrix of normal solutions in the lower half-plane, and 
$\rmX^+_p(z)$
as the matrix of normal solutions in the upper half-plane. Then, using the canonical embedding formula (\ref{eq:Can_solution_pole2}) we obtain:

\begin{equation}
\label{eq:plane_solution_pole2}
{\rmU}^{+}(z,z_i) = -\frac{{\rm X}_p^{+}(z)\left(\rmX^-_p(z_i)  - {\rm R}\right)^{-1}{\rm r}}{z-z_i}. 
\end{equation}
The latter is the plane wave embedding formula in the Fourier space.

\section{Embedding formula for canonical scattering problems}
\label{sec:canonical}
We will illustrate the idea of embedding on the simplest diffraction problems on a lattice, where all results can be checked directly. Below we avoid lengthy derivation and present only final expressions for embedding formulae. For the derivation of the Wiener--Hopf equations for discrete diffraction problems the reader is referred to \cite{Korolkov2025}. The embedding formula itself can be derived by mimicking the derivations from \cite{Korolkov2024}.

\subsection{The half-plane problem}
\label{sec:half_plane}
\begin{figure}[ht]
 	\centering{
 	\includegraphics[width=0.5\textwidth]{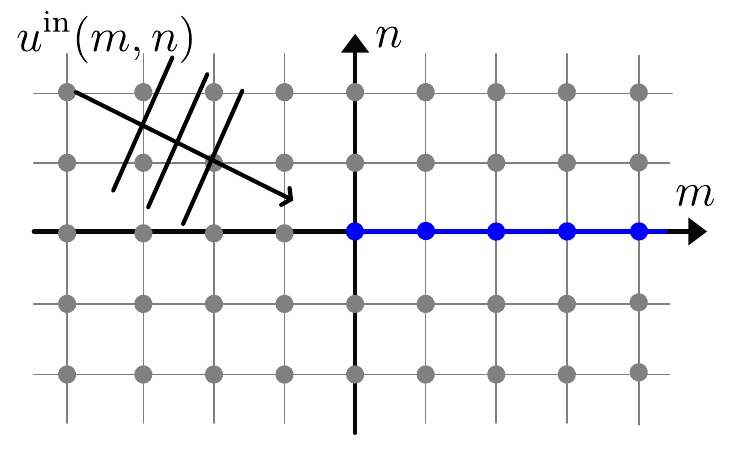}
 	}
 	\caption{Geometry of the problem of diffraction by a half-plane}
 	\label{fig: half-plane discrete}
 \end{figure}

The geometry of the problem is shown in Figure~\ref{fig: half-plane discrete}. 
Let the Dirichlet boundary condition be satisfied on {the positive} half-line,
\begin{equation}
u(m,0) = 0, \quad m\geq0,
\end{equation}
Let the total field satisfy the Helmholtz equation~(\ref{eq:Helmholtz}) in the rest of the domain and be represented as the sum of an incident plane wave and a scattered field:
\begin{equation}
\label{eq:inc_wave}
u(m,n) = u^{\rm in}(m,n) + u^{\rm sc}(m,n), \quad u^{\rm in}(m,n) = (\lx^{\rm in})^{-m} (\ly^{\rm in})^{-n},
\end{equation} 

The scattered field satisfies the radiation condition in terms of the limiting absorption principle due to a small imaginary part to the wavenumber \(\lk\). Moreover, consider a second plane wave problem with different incidence parameter $\beta_1^{\rm in}$, the incident wave $u_1^{\rm in}(m,n) = (\lx_1^{\rm in})^{-m} (\ly_1^{\rm in})^{-n}$. For both incident plane waves we can define directivities $S(\beta,\beta^{\rm in})$ and $S(\beta,\beta_1^{\rm in})$, as in Section~\ref{subsec:directivity}. Next we will derive the embedding formula that will link these directivities. 


Using the Wiener--Hopf formulation from \cite{Korolkov2025}  and following the method on normal solutions outlined above, we obtain the embedding formula in the Fourier space:
\begin{equation}
\label{hp_embed}
\Psi^+(s,s^{\rm in}) = -\frac{X^+(s,s^{\rm in}_1)(X^-(s^{\rm in},s_1^{\rm in}))^{-1}}{1-s(s^{\rm in})^{-1}},
\end{equation}
where
\[
X^+(s,s^{\rm in}_1) = (1-s(s^{\rm in}_1)^{-1})\Psi^+(s,s^{\rm in}_1),\quad  X^-(s,s^{\rm in}_1) = (1-s(s^{\rm in}_1)^{-1})\Psi^-(s,s^{\rm in}_1) - 1,
\]
\[
\Psi^+(\lx,s^{\rm in}) = \sum_{m=0}^{\infty}\lx^{m}\ptl_{(m,0)}[u^{\rm sc}],\quad \Psi^+(\lx,s^{\rm in}_1) = \sum_{m=0}^{\infty}\lx^{m}\ptl_{(m,0)}[u_1^{\rm sc}],
\]
\[
\Psi^-(\lx,s^{\rm in}_1) = \sum_{m=-\infty}^{-1}\lx^{m}u_1^{\rm sc}(m,0),
\]
\[
\ptl_{(m,0)}[u] = 1/2(u(m-1,0)+u(m+1,0)+(\lk^2-4)u(m,0)) + u(m,1).
\]
Here we have introduced a discrete analogue of a normal derivative $\ptl_{(m,0)}[\cdot]$. A detailed discussion of the definition can be found in \cite{Korolkov2025}, and the full definition is given in Appendix~\ref{app:reciprocity}

Using the equation (\ref{eq:directivity_link}) that links a discrete Fourier transform on the surface of the scatterer with a directivity of the scattered field, and the symmetry relation
\[
(X^-(s,s_1^{\rm in}))^{-1} = -\frac{X^+(s^{-1},s_1^{\rm in})}{X^+((s_1^{\rm in})^{-1},s_1^{\rm in})},
\]
the embedding formula can be expressed in terms of directivities:
\begin{equation}
\label{eq:embedding_half_plane}
    S\left(\beta,\beta^{\rm in}\right) = \frac{\hat S\left(\beta,\beta_1^{\rm in}\right)\hat S\left(\beta^{\rm in},\beta_1^{\rm in}\right)}{(1-(s_{\beta}s^{\rm in})^{-1})\hat S\left(\beta_1^{\rm in},\beta_1^{\rm in}\right)},
\end{equation}  
where \(s_{\beta}\) is defined in Remark~\ref{rem:label2}, and we have introduced the notation:
\[
\hat S\left(\beta,\beta^{\rm in}\right) = (1-(s_{\beta}s^{\rm in})^{-1})S\left(\beta,\beta^{\rm in}\right).
\]
Equation (\ref{eq:embedding_half_plane}) is a lattice analogue of the plane wave embedding formula for the directivities \cite{Biggs2006}. For the sake of completeness we derive the edge Green's version of the embedding formula in Appendix~\ref{app:Embedding_Edge}.

\subsection{The strip problem}
\label{sec:strip}
\begin{figure}[ht]
    \centering
    \includegraphics[width=0.5\linewidth]{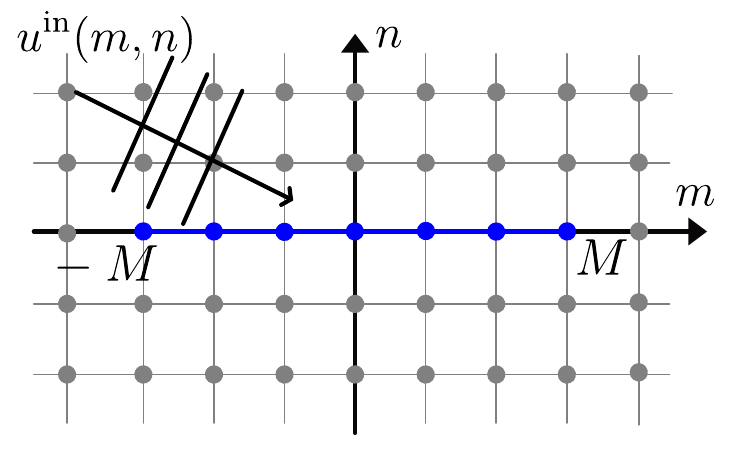}
    \caption{Geometry of the problem of diffraction by a finite strip}
    \label{fig:strip_geometry}
\end{figure}

A more complicated example is the problem of diffraction by a finite strip. The geometry of the problem is shown in Figure~\ref{fig:strip_geometry}. Let the total field be presented as a sum of the incident and scattered field (\ref{eq:inc_wave}),  satisfy Dirichlet boundary conditions on the strip,
\[
u(m,0) = 0,\quad  M\geq m\geq -M,
\]
satisfy the Helmholtz equation (\ref{eq:Helmholtz}) in the rest of the domain, and let the scattered field satisfy the radiation condition.

 Consider two additional plane wave problems with incidence parameters $\beta_1,\beta_2$ such that $\beta_2 = -\beta_1$. From the symmetry of the problem it follows that
\[
S(\beta,\beta_2) = S(-\beta,\beta_1).
\]

Similarly to the half-plane problem, using the Wiener--Hopf formulation from \cite{Korolkov2025}, and following the method of normal solutions, one can obtain the embedding formula:
\[
S(\beta,\beta^{\rm in}) = \frac{1}{1-(s_\beta s^{\rm in})^{-1}}\left(\frac{\hat S\left(\beta^{\rm in},\beta^{\rm in}_1\right)}{\hat S\left(\beta_1^{\rm in},\beta^{\rm in}_1\right)}\hat S\left(\beta,\beta^{\rm in}_1\right) +\frac{\hat S\left(\beta^{\rm in},\beta^{\rm in}_2\right)}{\hat S\left(\beta_2^{\rm in},\beta^{\rm in}_2\right)} \hat S\left(\beta,\beta^{\rm in}_2\right)\right).
\]

\subsection{The wedge problem}
\label{sec:wedge}
\begin{figure}[ht]
    \centering
    \includegraphics[width=0.5\linewidth]{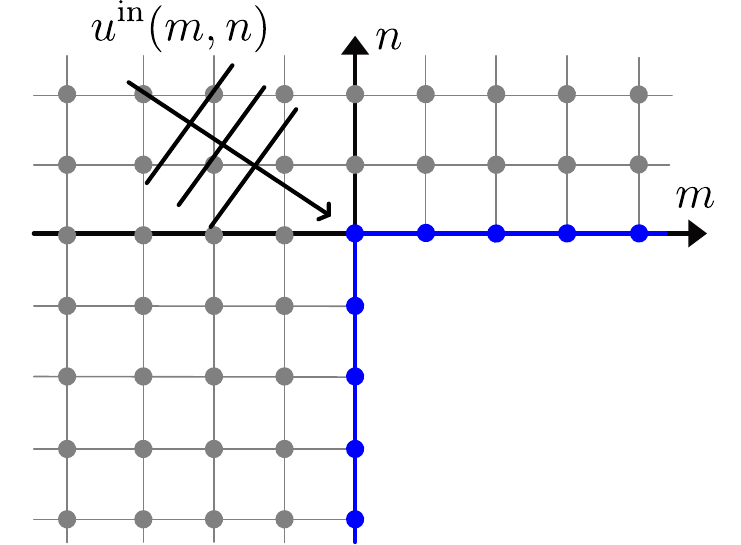}
    \caption{Geometry of the problem of diffraction by a right-angled wedge}
    \label{fig:wedge_geom}
\end{figure}

Now consider a problem of diffraction by a right-angled wedge. The geometry of the problem is shown in Figure~\ref{fig:wedge_geom}. Let the total field be presented as a sum of the incident and scattered field (\ref{eq:inc_wave}),  satisfy Dirichlet boundary condition on the wedge,
\[
u(m,0) = 0,\quad  m\geq 0,\quad u(0,n) = 0,\quad  n\leq 0,
\]
satisfy the Helmholtz equation (\ref{eq:Helmholtz}) in the rest of the domain, and let the scattered field satisfy the radiation condition.

Consider two additional plane wave problems with incident parameters $\beta_1,\beta_2$ such that  $\beta^{\rm in}_1 = 1/\beta^{\rm in}_2$. Then, from the symmetry of the problem we have:
\[
S(\beta,\beta^{\rm in}_2) = S(1/\beta,\beta^{\rm in}_1).
\]

Again, using the Wiener--Hopf formulation from \cite{Korolkov2025}, and following the method of normal solutions, one can obtain the embedding formula:
\begin{equation}
\label{eq:embeding_wedge}
S(\beta,\beta^{\rm in}) = \frac{\tilde S(\beta,\beta^{\rm in}_2)\tilde S(\beta^{\rm in},\beta^{\rm in}_1)-\tilde S(\beta,\beta^{\rm in}_1)\tilde S(\beta^{\rm in},\beta^{\rm in}_2)}{(s_\beta+ s_\beta^{-1} - s^{\rm in}-(s^{\rm in})^{-1})\tilde S(\beta^{\rm in}_1,\beta^{\rm in}_2)},
\end{equation}
where we used notation (\ref{eq:Mod_Dir_wedge})

It may seem surprising that the term in the dominator is not just a simple pole $s_\beta=s^{\rm in}$ like it was in the previous two examples. However, there is a clear explanation to this both from the Wiener--Hopf and operator approach (which is formulated in the next section) to embedding. Indeed, there is a preparatory work that needs to be done before the Wiener--Hopf problem can be formulated, since the  boundary orthogonal to  the $m$-axis is involved, and thus will lead to a spectral term with complicated domain of analyticity in $s$-variable. One way to deal with it is to apply the method of reflections to the half-line $m=0$, $n<0$, and in the result consider a diffraction problem on a manifold with boundaries lying on $m$-axis. Wiener--Hopf problem then formulated in a standard way, and  embedding formula is derived using a generalized version of (\ref{eq:plane_solution_pole2}). The derivation is similar to the continuos case that was studied in details in \cite{Korolkov2024}. Note, after the application of the method of reflections  the forcing term will contain two poles $s_\beta=s^{\rm in}$  and $s_\beta=1/s^{\rm in}$ which explains the denominator of (\ref{eq:embeding_wedge}). 

The operator explanation follow from the fact that the embedding operator needs to preserve boundary conditions both on line $m=0$ and line $n=0$, thus it should differ from the embedding operator for the half-plane and strip problems. We give the explicit expression for the embedding operators in the next section.


\section{Embedding formula in the general case}
\label{sec:embed_gen}
\begin{figure}
    \centering
    \includegraphics[width=0.53\linewidth]{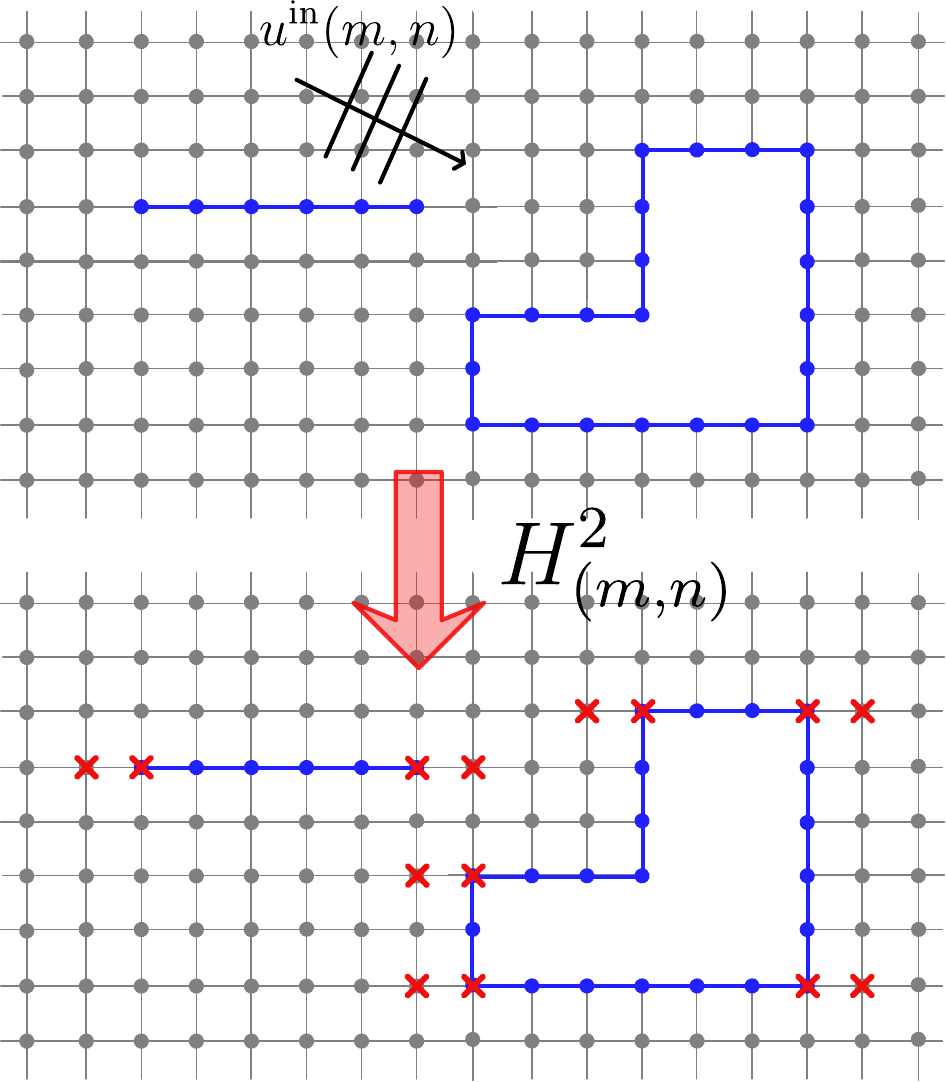}
    \caption{An example of a boundary value problem for lattice Helmholtz equation}
    \label{fig:general_case}
\end{figure}

Once the only two canonical geometries (the half-plane and the wedge) of the square lattice are studied it is easy to see how to derive the embedding formula in a general case. Indeed, let us consider a diffraction problem by several obstacles of  arbitrary shape. An example is shown in Figure~\ref{fig:general_case}, up. As previously, let the total field be presented as a sum of the incident and scattered field (\ref{eq:inc_wave}), satisfy Dirichlet boundary conditions on scatterers,
satisfy the Helmholtz equation (\ref{eq:Helmholtz}) in the rest of the domain, and let the scattered field satisfy the radiation condition.  

To formulate the embedding formula using the Weiner--Hopf formulation one needs to reduce the original problem to a Weiner--Hopf equation first. It can be done in the same manner as it was done for the wedge problem in \cite{Korolkov2025}, but it leads to lengthy expressions and cumbersome computations. In order to keep things simple we will follow an operator approach to embedding. A lattice version of the algorithm proposed in \cite{Craster2003} and modified in \cite{Biggs2006} can be summarised as follows:
\begin{enumerate}
 \item Find operator, \(H\), so that \(H [u^{\rm in}]=0\), where \(u^{\rm in}\) is the incident plane wave, and for total field \(u\), \(H [u]\) is still a solution to the Helmholtz equation  with the same boundary conditions  except at finite number of nodes, where $H[u]$ has point sources.
 \item Introduce edge Green's functions $v_i$ that are solutions to the Helmholtz equation with the boundary conditions, but have point sources that correspond to point sources of $H[u]$.
 \item Show that \(L[u,v_i]\equiv H [u]+\sum_i K_i v_i\) satisfies the Helmholtz equation, boundary conditions, and radiation condition, with some constants  $K_i$ that are to be determined. Since  \(L[u,v_i]\) is the solution of the diffraction problem, then, by uniqueness \(L[u,v_i]=0\), giving the weak form of embedding.
 \item Using the uniqueness argument once more, express edge Green's functions in terms of the plane wave solutions.
 \item  Take the far-field limit and use reciprocity to obtain the strong embedding formula.
 \end{enumerate}

The weak point of this algorithm is that it starts from a heuristic first step where an appropriate operator should be guessed. However, we can bypass it by using the embedding formula defined via Wiener--Hopf perspective. Indeed, the denominators of (\ref{eq:embedding_half_plane}) and (\ref{eq:embeding_wedge}) are the Fourier transforms of the corresponding operators. Thus, the embedding operator for the half-plane and the strip problems is
\[
H^1_{(m,n)}[u]=u(m,n) - (s^{\rm in})^{-1}u(m-1,n).
\]
For the half-plane, direct check shows that it annihilates the incident wave, satisfies Helmholtz equation everywhere outside the scatterer, and preserves the boundary condition everywhere except in the origin $(0,0)$. For the strip problem it will not satisfy the Dirichlet boundary condition in $(-N,0)$, and the Helmholtz equation in $(N+1,0)$. For the wedge problem, the embedding operator has the following form:
\[
H^2_{(m,n)}[u] = u(m+1,n) + u(m-1,n) - \left((s^{\rm in})+(s^{\rm in})^{-1}\right)u(m,n).
\]
The definition of $H^2_{(m,n)}$ should be clarified, since for the wedge problem its value is undefined on half-line  $m=0$, $n<0$. We define it there by assigning the field values according to the method of reflections: 
\[
u(1,n) \equiv - u(-1,n),\quad n<0. 
\]
In the result $H^2_{(m,n)}[u]$ satisfies Helmholtz equation on half-line $m=-1$, $n<0$ and Dirichlet boundary conditions on half-line $m=0$, $n<0$. Thus, $H^2_{(m,n)}[u]$ does not satisfy the Helmholtz equations with boundary conditions only in two points $(-1,0)$ and $(0,0)$.

It worth mentioning that similarly to the continuous diffraction problem \cite{Craster2003}, embedding operator is of the first order for the strip problem, and of the second order for the wedge problem. 

Then, by following  the above algorithm further we obtain the embedding formula. Namely,  $H^2_{(m,n)}[u]$ satisfies the Helmholtz equation with boundary conditions everywhere except the vertices corresponding to edges and outer right-angle wedges and in the points neighbouring to these vertices. For example, for the geometry shown in Figure~\ref{fig:general_case}, up, these points are shown by crosses in Figure~\ref{fig:general_case}, down. Thus, $H^2_{m,n}[u]$ can be expressed as a combination of edge Green's functions: 
\[
H^2_{(m,n)}[u] = \sum_{l=1}^N K_l v_l(m,n),
\]
where $K_l$ are some constants, \(N\) is the number of points where Helmholtz equation is not satisfied, and $v_l$ are solutions of the following inhomogeneous Helmholtz equation:
\[
\Delta_{(m,n)}[v_l] + \lk^2 v_l(m,n) = C_l\delta_{m-m_l}\delta_{n-n_l},
\]
where $(m_l,n_l)$ are coordinates of the source that is placed in one of the points where the Helmholtz equation or boundary conditions are not satisfied, and $C_l$ are some constants. Additionally, $v_l(m,n)$ should satisfy the boundary and radiation conditions. Then, using the uniqueness argument once again we express the edge Green's functions in terms of the plane wave solutions. Consider $N$ plane wave solutions $H[u_l]$, $l=1,\ldots,N$, and by choosing $l$ coefficients $A_l$ and  we arrive to the weak embedding formula for fields: 
\begin{equation}
\label{eq:weak_embeding_field}
H^2_{(m,n)}[u] = \sum_{l=1}^NA_lH^2_{(m,n)}[u_l].
\end{equation}
Note, that coefficients $A_l$ depend on the angle of incidence $\theta^{\rm in}$.
Using the far-field representation and taking the far-field limit, we obtain
\begin{equation}
\label{eq:weak_embeding}
\tilde S(\beta,\beta^{\rm in}) = \sum_{l=1}^NA_l\tilde S(\beta,\beta_l^{\rm in}),
\end{equation}
where we used notation (\ref{eq:Mod_Dir_wedge}).
The latter is known as the weak embedding formula for directivities. To determine unknown coefficients $A_l$ let us make use of reciprocity theorem (\ref{eq:recipr_dir}),
and hence
\[
\tilde S(\beta,\beta^{\rm in}) = -\tilde S(\beta^{\rm in},\beta).
\]
Thus, the coefficients are determined from the following system of linear equations:
\begin{equation}
\label{eq:srong_der_embeding}
-\tilde S(\beta^{\rm in},\beta^{\rm in}_p) =  \sum_{l=1}^NA_l\tilde S(\beta^{\rm in}_p,\beta_l^{\rm in}), \quad p = 1\ldots N.
\end{equation}
Solving the latter and substituting it into ({\ref{eq:weak_embeding}}) we obtain the strong embedding formula.

\begin{rem}
In particular, it means that \(N\) values \(\tilde S(\beta^{\rm in},\beta^{\rm in}_p)\) for \(p = 1\ldots N\) determine the function  \(\tilde S(\beta,\beta^{\rm in})\). Also if we do not know the value of \(N\) then it is possible to find it by considering the rank of the \(M \times M\) matrix composed of elements  \(S(\beta_p^{\rm in},\beta^{\rm in}_l)\) when \(M\) is taken sufficiently large so that \(M>N\), see~Figure~\ref{fig:ranks}.
    
\end{rem}

\section{Numerical illustrations}
\label{sec:numerics}
A standard computational approach to the solution of outer scattering problems in continuos case is the method of boundary integral equation. It was shown in \cite{Martinsson2009, PobletPuig2018} that an analogue can be build on lattices, which is known as the method of boundary algebraic equations (BAE). Below we formulate it in a slightly different way using Green's identity on lattices, see Appendix~\ref{app:reciprocity} and \cite{Korolkov2025}.

Consider a problem of diffraction by a compact scatterer. Denote the domain outside of the scatterer by $\Omega$, and  $\ptl \Omega$ as the boundary of the domain. Now apply Green's identity (see Appendix~\ref{app:reciprocity}) in $\Omega$ with $u^{\rm sc}_\bnu$ and $G^{\bmu}_\bnu$ where the latter is the free field Green's function (see Appendix~\ref{app:app_Green}). We obtain:
\[
\sum_{\bnu \in \ptl\Omega}(\ptl_\bnu[u^{\rm sc}]G^\bmu_\bnu - \ptl_{\bnu}[G^\bmu]u^{\rm sc}_{\bnu}) = 0, 
\]
where $\bnu$ is a multi-index $\bnu = \{m,n\}$.
Using the fact that $u^{\rm sc}_{\bnu} =-u^{\rm in}_{\bnu}$ on the surface of the scatterer and $G^\bmu_\bnu$ can be calculated via the integral (\ref{eq:Greens_integral}) we arrive to the following system of algebraic equations:
\begin{equation}
\label{eq:BAE}
\sum_{\bnu \in \ptl\Omega}K_{\bmu\bnu}\ptl_\bnu[u^{\rm sc}] = F_\bmu,\quad \bmu \in \ptl\Omega,
\end{equation}
where 
\[
K_{\bmu\bnu} = G^\bmu_\bnu ,\quad F_\bmu = - \sum_{\bnu\in\ptl\Omega}\ptl_{\bnu}[G^\bmu]u^{\rm in}_{\bnu}.
\]
Solving it with respect to $\ptl_\bnu[u^{\rm sc}]$ and using (\ref{eq:dir_gen}) we recover the directivity of the scattered field.

Note that the scattered field can be reconstructed everywhere using Green's identity:
\begin{equation}
\label{eq:rec_formula}
u^{\rm sc}_{\bmu} = \sum_{\bnu \in \ptl\Omega}(\ptl_\bnu[u^{\rm sc}]G^\bmu_\bnu + \ptl_{\bnu}[G^\bmu]u^{\rm in}_{\bnu}), \quad \bnu \in \Omega.
\end{equation}

We have developed a finite element based solver (FEM) that solves (\ref{eq:BAE}) and calculates (\ref{eq:rec_formula}). The solver was implemented in MATLAB environment, and designed in  such a way that it can handle finite obstacle of any shape, and any 9 point approximation of the Laplace operator. The solver is publicly available on GitHub \cite{MWM_BAE_2D_2025}.

To validate the embedding formula numerically  we consider several diffraction problems by obstacles with different geometry. 

\begin{figure}
    \centering
    \includegraphics[width=0.49\linewidth]{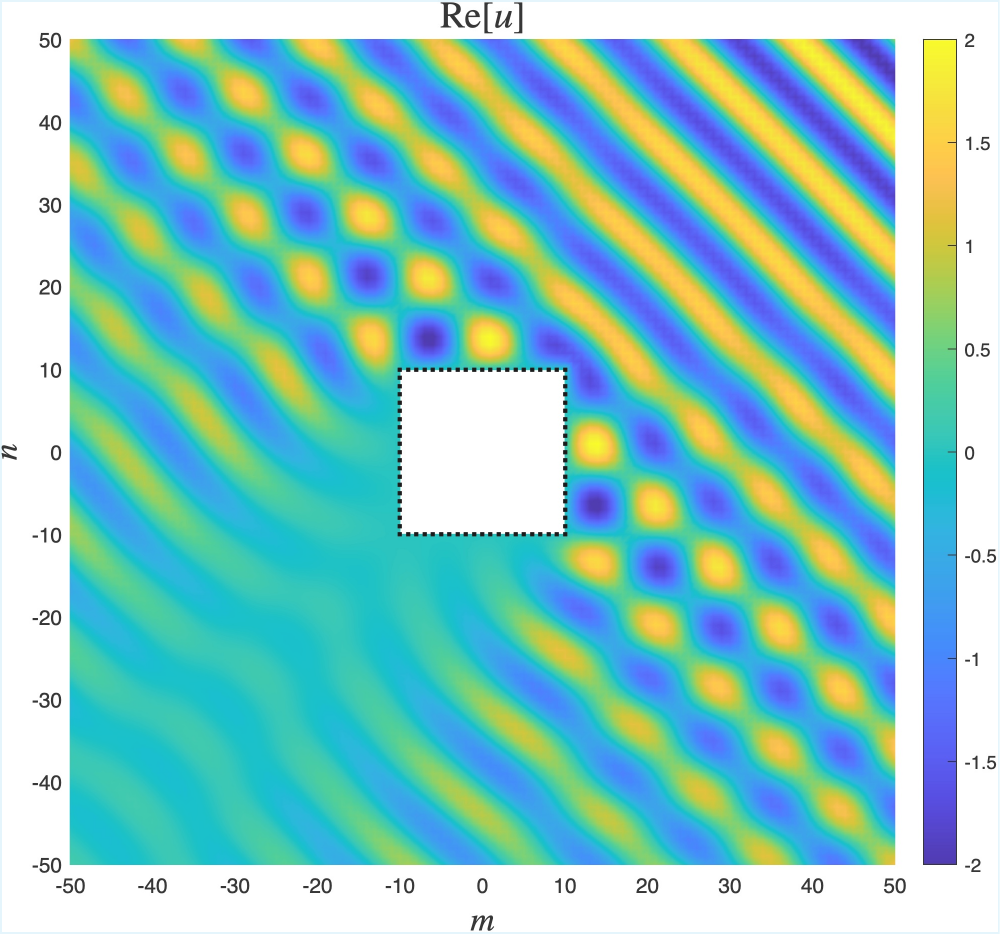}
    \includegraphics[width=0.49\linewidth]{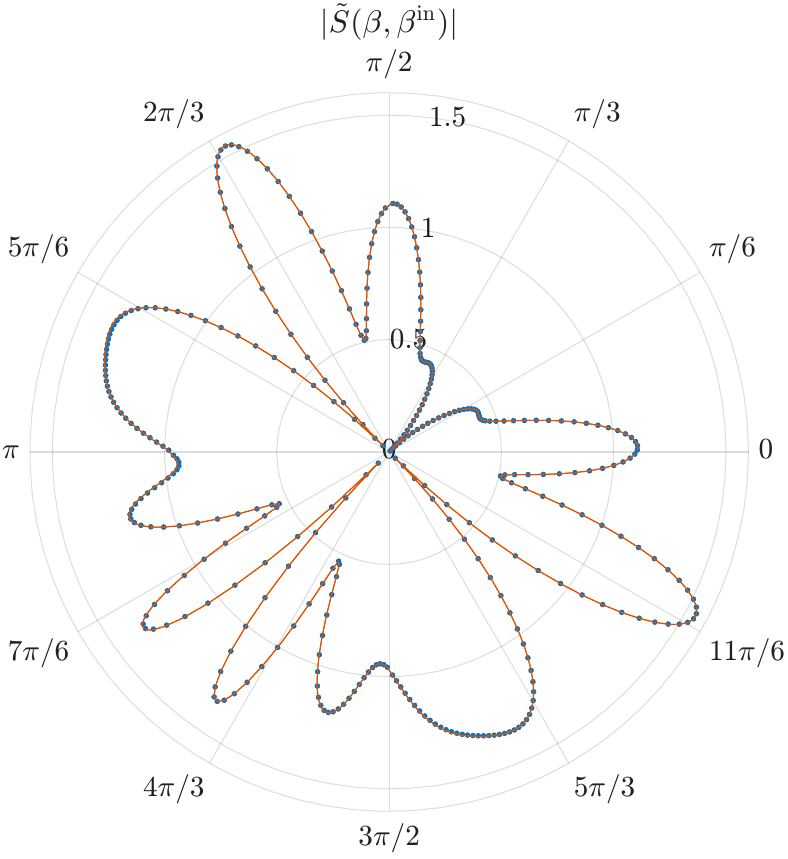}
    \caption{Real part of the total field diffracted by a square (left), and absolute value of the modified directivity calculated directly (solid line) and via the embedding formula (dotted line) (right)}
    \label{fig:square_numerics}
\end{figure}

First, consider the problem of diffraction by  a square of side that consists of  $L$ nodes. Real part of the total field $u(m,n)$ obtained directly from (\ref{eq:BAE}) and reconstruction formula (\ref{eq:rec_formula}) is shown in Figure~\ref{fig:square_numerics}, left. 
Here, $L=21$, $\lk=0.6 + 0.01i$, and $\beta^{\rm in} = 1$, which corresponds to angle of incidence $\theta^{\rm in} = \pi/4$ if measured from $n$-axis. As it was shown in Section~\ref{fig:general_case} embedding formula requires $8$ independent plane wave solutions in this case. We chose the following incidence parameters:                     $\beta^{\rm in}_1 = 0.2769$, $\beta^{\rm in}_2 = 0.4710$, $\beta^{\rm in}_3 = 0.6994$, $\beta^{\rm in}_4 = 0.9900$, $\beta^{\rm in}_5 = 11.3999$, $\beta^{\rm in}_6 = 2.0691$, $\beta^{\rm in}_7 = 3.4763$, $\beta^{\rm in}_8=9.0542$.  
Absolute value of the modified directivity $\tilde S(\beta,\beta^{\rm in})$ is presented in Figure~\ref{fig:square_numerics}, right. Solid lines were obtained directly using (\ref{eq:directivity_link}), and  dotted lines were obtained using the embedding formula (\ref{eq:weak_embeding}). For general numerically robust method on how to recover the directivity from the modified directivity see~\cite{Gibbs_2018}.

\begin{figure}
    \centering
    \includegraphics[width=0.49\linewidth]{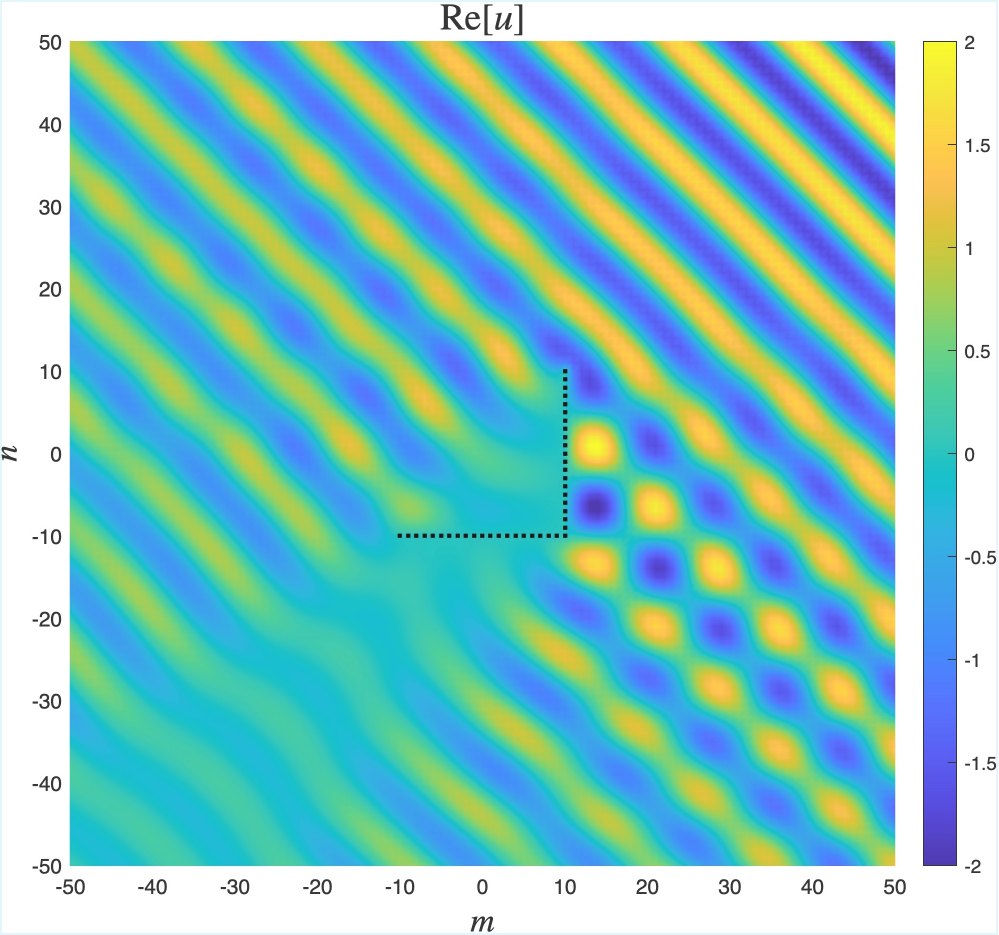}
    \includegraphics[width=0.49\linewidth]{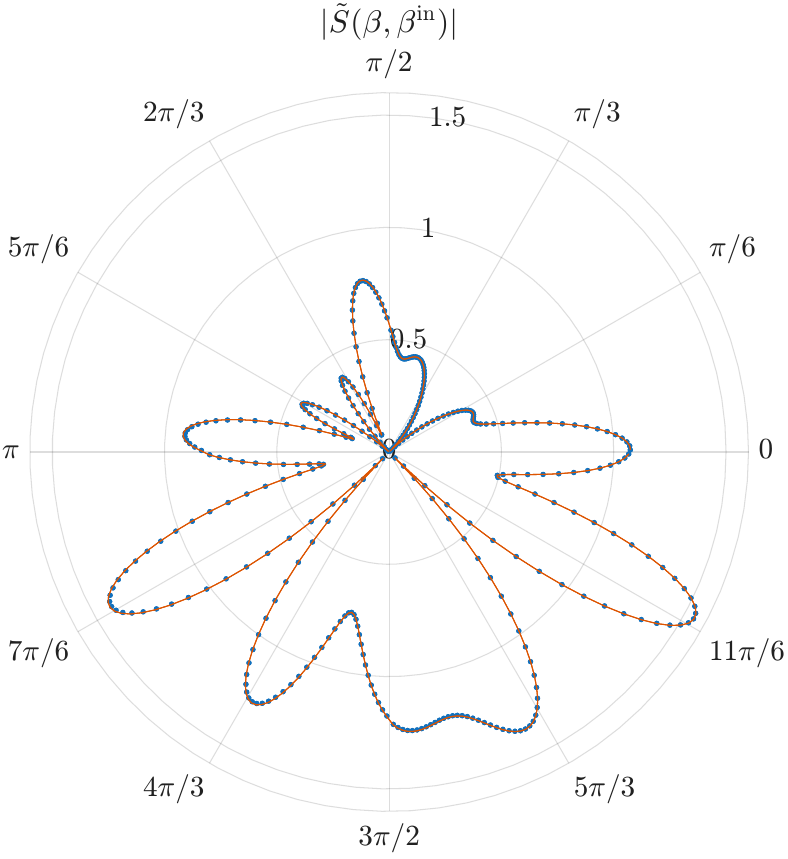}
    \caption{Real part of the total field diffracted by a right angle (left), and absolute value of the modified directivity calculated directly (solid line) and via the embedding formula (dotted line)}
    \label{fig:angle_numerics}
\end{figure}

Second, consider a problem of diffraction by a finite right angle. The geometry of the obstacle and the real part of the total field is shown in Figure~\ref{fig:angle_numerics}, left. Each side of the angle consists of $21$ nodes, $\lk=0.6 + 0.01i$, and $\beta^{\rm in} = 1$. Embedding in this case requires $6$ independent solutions. We chose the following incidence parameters:                    
$\beta^{\rm in}_1 = 0.3390$, $\beta^{\rm in}_2 = 0.6181$, $\beta^{\rm in}_3 = 0.9900$, $\beta^{\rm in}_4 = 1.5823$, $\beta^{\rm in}_5 = 2.8560$, $\beta^{\rm in}_6 = 9.0542$. This example demonstrates that the inner angle of the obstacle do not increase the number of terms \(N\) in the embedding formula.
Absolute value of the modified directivity $\tilde S(\beta,\beta^{\rm in})$ is presented in Figure~\ref{fig:angle_numerics}, right.

\begin{figure}
    \centering
    \includegraphics[width=0.49\linewidth]{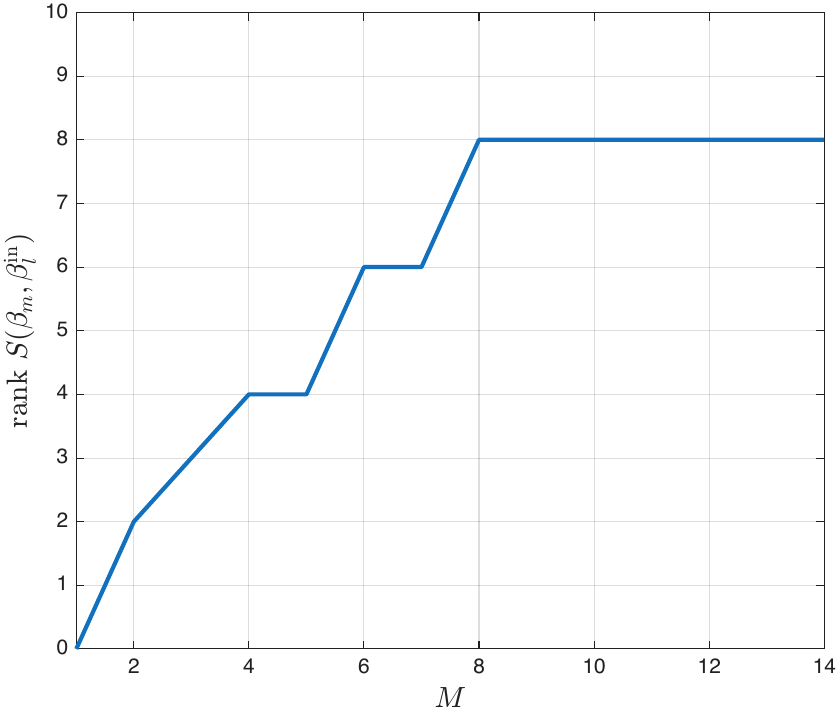}
    \includegraphics[width=0.49\linewidth]{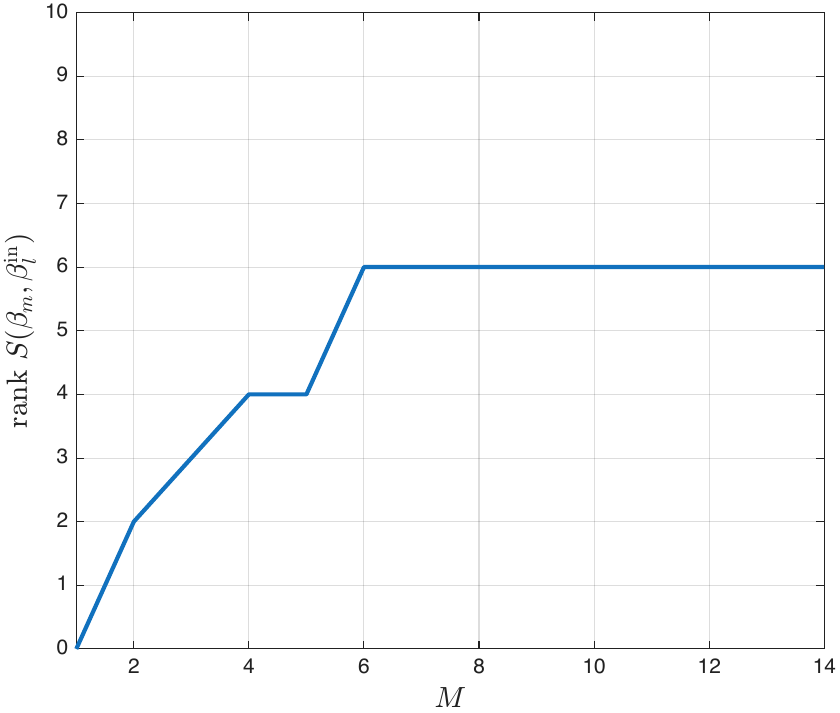}
    \caption{Ranks of matrix composed of \(\tilde S(\beta_m,\beta_l^{\rm in})\) computed via the singular value decomposition for the problem of diffraction by a square (left) and the problem of diffraction by a right angle (right)}
    \label{fig:ranks}
\end{figure}

Let us now demonstrate that \(N\) can be recovered from matrix composed of 
\(\tilde S(\beta_m,\beta_l^{\rm in})\), provided these values are known for 
\(m,l = 1,2,\dots,M\) with \(M\) sufficiently large, by computing the rank of this matrix. 
A numerically reliable approach is to perform a singular value decomposition and count 
the number of singular values exceeding a chosen threshold. In the examples below, the 
threshold was set to \(5 \times 10^{-5}\). The results of this computation for several values of \(M\) are shown in 
Figure~\ref{fig:ranks} for the diffraction problems involving a square (left) and a 
right angle (right). As expected, once \(M\) reaches \(N\), the rank of the matrix 
composed of \(\tilde S(\beta_m,\beta_l^{\rm in})\) stabilises and remains constant.
\section{Conclusion}
\label{sec:conclusion}
In this article, we derived embedding formulae for an arbitrary shaped Dirichlet obstacle on  a square lattice.  This was done by using the Wiener--Hopf perspective to embedding to tackle some simple problems (half-plane, strip, wedge). Then, from the structure of the solutions the embedding operator was deduced, and operator approach was used to obtain the general embedding formula.  The latter makes the current study self-contained and algorithmic. One of ours motivations was to portray the power of Wiener--Hopf approach to embedding, and how it effortlessly allows us to drive the embedding formula in any context where the problem can be formulated as a Wiener–Hopf equation. This means that it is easily  extendable to other boundary conditions and other lattices. In fact, embedding formula can easily be obtained for any other cases where Wiener--Hopf is used, like for example~\cite{Mishuris_beam}.

It also worth to mention that results of this paper naturally continue the idea of analogy between continuos and lattice problems that was introduced in \cite{Korolkov2025} by extending it to the analogy between embedding formulae.

To demonstrate the validity of embedding formula numerically, a FEM based numerical solver was developed by authors \cite{MWM_BAE_2D_2025}.

Returning to the motivations stated in the introduction, our results deliver on the three main advantages of embedding for discrete diffraction:
\begin{enumerate}
\item  Computational efficiency. Only a finite number $N$ of auxiliary directivities, where $N$ is twice the number of corner points, must be computed. All remaining incident directions follow from the embedding formula without resolving boundary value problems.
\item Reconstruction from sparse data. If the directivity is known for a small set of incidence parameters, the full angular dependence can be reconstructed exactly. This is particularly valuable for numerics, inverse problems, and experimental contexts where only limited angular sampling is available.
\item Extracting geometric information. When $N$ is unknown, it can be recovered from the rank of a finite matrix of measured (or computed) directivities. Hence, the embedding framework provides a diagnostic tool for identifying the number of geometric features responsible for scattering.
\end{enumerate}

There are several natural directions for future work. First, although the number of required auxiliary incidence angles is dictated by geometry, the choice of these angles remains flexible. A systematic strategy for selecting optimal incidence parameters that balances conditioning, numerical stability, and physical interpretability would improve the robustness of the method, especially for complex geometries. Second, the Wiener–Hopf perspective suggests straightforward extensions to other boundary conditions (Neumann, impedance) and to more general lattices or multi‑point discretizations of Laplace operator. Finally, the operator interpretation of embedding opens the door to applications beyond scattering, such as cloaking, defect design, and active control, where tailored Green’s functions may serve as a more convenient basis than plane waves.

\section*{Acknowledgment}
 A.V.K. is supported by a Royal Society
Dorothy Hodgkin Research Fellowship which also supported A.I.K via the Royal Society Research Fellows Enhanced Research Expenses.
Authors gratefully acknowledge the support of the EU H2020 grant MSCA-RISE-2020-101008140-EffectFact. The authors would also like to thank the Isaac Newton Institute for Mathematical Sciences (INI) for their support and hospitality during the programme ``WHT Follow on: the applications, generalisation and implementation of the Wiener--Hopf Method''(WHTW02), where work on this paper was undertaken and supported by EPSRC grant no EP/R014604/1. 

\bibliographystyle{unsrt}
\bibliography{Embedding_lattices,Embedding}

\appendix
\appendixpage
\section{Green's identity. Reciprocity relation}
\label{app:reciprocity}
An important step of derivation of BAE is based on the application Green's identity. It is well known for continuos problems, but is not often used on lattices. We summarise here, for more details see \cite{Korolkov2025}. Consider two solutions of the lattice Helmholtz equation: 
\[
\Delta_{\bnu}[u] + k^2 u_{\bnu} = f_{\bnu},
\quad
\Delta_{\bnu}[w] + k^2 w_{\bnu} = g_{\bnu}, \quad \bnu = {(m,n)}.
\]
Then, in a  domain $\Omega$ with the boundary $\ptl \Omega$ the following relation (Green's identity) is satisfied:
\begin{equation}
\label{eq:Green's_lattice}
\sum_{{\bnu} \in \ptl \Omega} \left(\ptl_{\bnu}[u] w_{\bnu}  - \ptl_{\bnu}[w] u_{\bnu}\right) =    \sum_{{\bnu} \in\Omega}\left(g_{\bnu} u_{\bnu} - f_{\bnu} w_{\bnu}\right).
\end{equation}
Here, $\ptl_{\bnu}$ is  a discrete analogue of the normal derivative for the boundary nodes:
\begin{equation}
\label{eq:disc_derivative}
\ptl_{\bnu}[u] = \sum_{{\bmu} \in \ptl \Omega\cup\Omega_{\rm adj}} \alpha_{{\bnu}{\bmu}} u_{\bmu} ,\quad {\bnu}\in \ptl\Omega,  
\end{equation}
where $\alpha_{{\bnu}{\bmu}}$ is defined differently for different parts of the boundary. 
\begin{figure}[h]
    \centering
    \includegraphics[width=0.4\linewidth]{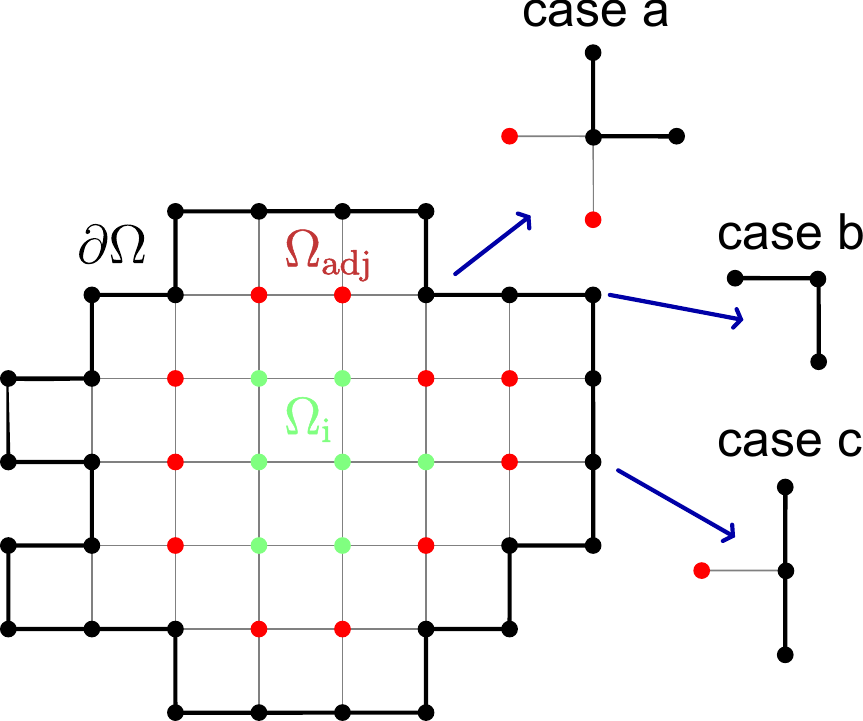}
    \caption{A general domain for Green's identity on lattices. Figure is adapted from \cite{Korolkov2025}}
    \label{fig:latticeGreensdomain}
\end{figure}
There are three possible cases (see Figure~\ref{fig:latticeGreensdomain}): 
\begin{itemize}
\item ${\bnu}$ lies on an external right angle, see Figure~\ref{fig:latticeGreensdomain}, case a. Then:
\[
\alpha_{{\bnu}{\bmu}}=\begin{cases}
1/2, & {\bnu}\neq {\bmu}, \text{ ${\bnu}$ is adjacent {to} ${\bmu}$ and ${\bmu} \in \ptl \Omega$  }\\
1, & {\bnu}\neq {\bmu}, \text{ ${\bnu}$ is adjacent {to} ${\bmu}$ and ${\bmu} \in \Omega_{\rm adj}$}\\
3(\lk^2/4 - 1), & {\bnu} = {\bmu},\\
0, & \text{otherwise}.
\end{cases}
\]
\item ${\bnu}$ lies on an internal right angle, see Figure~\ref{fig:latticeGreensdomain}, case b. Then:
\begin{equation}
\label{eq:lat_bound_Neum_b}
\alpha_{{\bnu}{\bmu}}=\begin{cases}
1/2, & {\bnu}\neq {\bmu}, \text{ ${\bnu}$ is adjacent {to} ${\bmu}$}\\
\lk^2/4- 1, & {\bnu} = {\bmu},\\
0, & \text{otherwise}.
\end{cases}
\end{equation}
\item ${\bnu}$ lies on a straight line, see Figure~\ref{fig:latticeGreensdomain}, case c. Then:
\[
\alpha_{{\bnu}{\bmu}}=\begin{cases}
1/2, & {\bnu}\neq {\bmu}, \text{ ${\bnu}$ is adjacent {to} ${\bmu}$ and ${\bmu} \in \ptl \Omega$  }\\
1, & {\bnu}\neq {\bmu}, \text{ ${\bnu}$ is adjacent {to} ${\bmu}$ and $ {\bmu} \in \Omega_{\rm adj}$}\\
2(\lk^2/4 - 1), & {\bnu} = {\bmu},\\
0, & \text{otherwise}.
\end{cases}
\]
\end{itemize}
We can easily deduce the reciprocity relation from the latter. Consider two solutions of the same boundary value problem with point source excitations located in two distinct points: 
\[
u_{\bnu}  \equiv v_{\bnu}^{\bmu_s},\quad  w_{\bnu}  \equiv v_{\bnu}^{\bmu_r},
\]
where $\bmu_s$ and $\bmu_r$ are indices of the point sources. Then, 
\[
v_{\bmu_s}^{\bmu_r} = v_{\bmu_r}^{\bmu_s}.
\]
Thus, we have for directivities
\[
S(\beta,\beta^{\rm in}) = S(\beta^{\rm in},\beta). 
\]

\section{Asymptotic estimation of the free field Green's function}
\label{app:app_Green}
Let us consider the Helmholtz equation on an infinite 2D lattice with a point source in the origin: 
\[
\Delta_\bnu[G] + \lk^2G_\bnu = \delta_{0,\bnu},
\]
where $\delta_{\bmu,\bnu}$ is Kronecker's symbol. Using Fourier transform the solution is expressed as follows:
\begin{equation}
\label{eq:Greens_integral}
G(m,n) = \frac{1}{2\pi i}\int_\sigma s^m q^n\frac{ds}{s(q-q^{-1})},
\end{equation}
where $\sigma$ is a unit circle bypassed in the positive direction. Let us rewrite the integral in the following form:
\[
G(m,n) = \frac{1}{2\pi i}\int_{\sigma}\exp\{N\phi(s,q)\}\frac{ds}{s(q-q^{-1})},
\]
where
\[
\phi(s,q) = \tilde m\log(s) + \tilde n\log(q),\quad N = \sqrt{m^2+n^2},\quad \tilde m = \frac{m}{N},\quad \tilde n = \frac{n}{N}.
\]
Let us then estimate it as $N\to \infty$. Saddle points $(s_\beta,q_\beta)$ of the equation are given by the following equation:
\begin{equation}
\label{eq:saddle_points}
\frac{m}{n} = \frac{s_\beta - s_\beta^{-1}}{q_\beta - q_\beta^{-1}}. 
\end{equation}
Detailed analysis shows that only one saddle point contribute to the integral. Note also that the saddle point $s_\beta$ is a function of $\beta = m/n$. Thus, using the saddle point method we obtain:
\begin{equation}
\label{eq:greens_asympt}
G(m,n) = \exp\{N\phi(s_\beta,q_\beta)-i\phi_0\}\sqrt{\frac{1}{2\pi N|\phi''(s_\beta)|}}\frac{1}{is_\beta(q_\beta-q_\beta^{-1})} + O\left(\frac{1}{N^{3/2}}\right),
\end{equation}
where
\[
\phi'' = \frac{d^2\phi}{ds^2}(s_\beta),\quad \phi_0 = \frac{\pi - {\rm arg}[\phi'']}{2}. 
\]
Denote 
\[
g(m,n) = \exp\{N\phi(s_\beta,q_\beta)-i\phi_0\}\sqrt{\frac{2\pi}{N\phi''(s_\beta)}}\frac{1}{s_\beta(q_\beta-q_\beta^{-1})},
\]
i.e. $g(m,n)$ is the leading term of the asymptotic expansion of $G(m,n)$. 
\section{Spectral presentation of the field in the general case. Link between the directivity and the spectrum}
\label{app:Directivity}
\begin{figure}
    \centering
    \includegraphics[width=.8\linewidth]{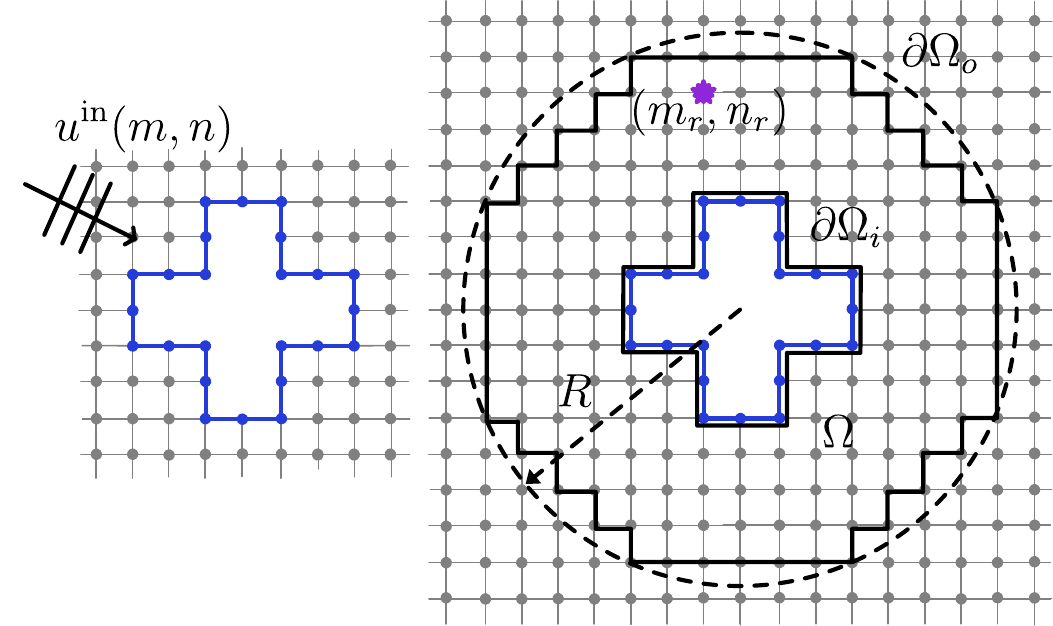}
    \caption{Sample geometry of the problem (left) and domain for Green's identity (right)}
    \label{fig:Greens_directivity}
\end{figure}

Consider an outer problem of diffraction by a compact scatterer.  Sample geometry is shown in Figure~\ref{fig:Greens_directivity}, left. 
Let the total field satisfy  the lattice Helmholtz equation (\ref{eq:Helmholtz}) and  Dirichlet boundary conditions on the surface of the scatterer. Apply Green's theorem to the domain $\Omega$ shown in Figure~\ref{fig:Greens_directivity}, right taking 
\[
u(m,n) = u^{\rm sc}(m,n),\quad w(m,n) = G(m_r-m,n_r-n)\equiv G_{\bnu_r}^{\bnu}, 
\]
where $\bnu_r = \{m_r,n_r\}$ are the indices of the source node.
Due to the radiation condition the integral along the big ``arc'' $\ptl \Omega_0$ tends to zero as $R\to\infty$. Thus, the Green's identity reduces to
\[
u^{\rm sc}(m_r,n_r) = \sum_{\bnu \in \ptl \Omega_i}(\ptl_\bnu[u^{\rm sc}]G^\bnu_{\bnu_r} - \ptl_\bnu[G_{\bnu_r}]u^{\rm sc}_\bnu),
\]
where $\ptl_\bnu$ is defined in (\ref{eq:disc_derivative}), $\ptl \Omega_i$ corresponds to the surface of the scatterer, see Figure~\ref{fig:Greens_directivity}.
Next we need to estimate the field as $\sqrt{m_r^2 + n_r^2}\to \infty$. Using the asymptotic representation 
(\ref{eq:greens_asympt}), we obtain the following:
\[
u^{\rm sc}(m_r,n_r) = g(m_r,n_r)\sum_{\bnu \in \Omega_i}\left(\ptl_{\bnu}[u^{\rm sc}]s_\beta^{-m}q_\beta^{-n}+ \ptl_{\bnu}[s_\beta^{-m}q_\beta^{-n}]u^{\rm in}_{\bnu}\right) + O((m^2_r + n^2_r)^{-3/2}),
\]
where $s_\beta,q_\beta$ are chosen in such a way that (\ref{eq:saddle_points}) is satisfied for $m=m_r$ and $n=n_r$. Finally, from the definition (\ref{eq:Far_field}) we obtain the following expression for the directivity of the scattered field:
\begin{equation}
S\left(\frac{m_r}{n_r}\right) = \sum_{\bnu \in \Omega_i}\left(\ptl_{\bnu}[u^{\rm sc}]s_\beta^{-m}q_\beta^{-n} + \ptl_{\bnu}[s_\beta^{-m}q_\beta^{-n}]u^{\rm in}_{\bnu}\right).
\label{eq:dir_gen}
\end{equation}
In the particular case of the half-plane problem (see \eqref{hp_embed} and below) we have:
\begin{equation}
\label{eq:directivity_link}
S\left(\beta\right) = \Psi^+(1/s_\beta,s^{\rm in}).
\end{equation}

\section{Embedding formula via edge Green's functions}
\label{app:Embedding_Edge}
Consider the problem of diffraction by a half-plane. Study the edge Green's function.
Let us introduce it as solution of the inhomogeneous Dirichlet problem:
\[
v(0,0) = C_1, \quad v(m,0) = 0, \quad m \geq -1,
\]
with Helmholtz equation 
\[
\Delta_{(m,n)}[v] + \lk^2 v(m,n) = 0
\]
satisfied in the rest of the domain. Here $C_1$ is some constant.
The problem can be equivalently formulated as a problem with a point source:
\[
\quad v(m,0) = 0, \quad m \geq -1,
\]
\[
\Delta_{(m,n)}[v] + \lk^2 v(m,n) = C_2\delta_{m,0}\delta_{n,0},
\]
where $C_2$ is some constant.
Let us show that 
\begin{equation}
\label{eq:C2_C1}
C_2=\sqrt{\eta_{o1}\eta_{o2}}C_1,
\end{equation}
where $\eta_{o1},\eta_{o2}$ are defined below. 

Indeed, this can be  done  using the Wiener--Hopf method which leads to the following equation:
\[
\Upsilon(s)V^-(s) - V^+(s)=0, 
\]
where
\[
V^-(s) = \sum_{m = -\infty}^{0}v(m,0)s^{m},\quad V^+(s) = \sum_{m=0}^\infty\ptl_{(m,0)}[v]s^m,
\]
\[
\Upsilon = (2\lx)^{-1}\sqrt{\left(\lx - \eta_{o1}\right)\left(\lx - \eta_{o2}\right)\left(\lx - \eta_{i1}\right)\left(\lx - \eta_{i2}\right)}.
\]
\[
\eta_{o1} = -\frac{d_1}{2} - \frac{i\sqrt{4-d_1^2}}{2}, \quad \eta_{i1}  = -\frac{d_1}{2} +\frac{i\sqrt{4-d_1^2}}{2}, \quad d_1 = \lk^2 - 2, 
\]
\[
\eta_{o2} = -\frac{d_2}{2} + \frac{\sqrt{d_2^2-4}}{2}, \quad \eta_{i2}  = -\frac{d_2}{2} -\frac{\sqrt{d_2^2-4}}{2}, \quad d_2 = \lk^2 - 6. 
\]
The solution is given by 
\[
V^-(s) = \frac{2sA}{\sqrt{(s-\eta_{i1})(s-\eta_{i2})}},\quad V^+(s) = A\sqrt{(s-\eta_{o1})(s-\eta_{o2})}.
\]
where the constant $A$ is determined from the field value in the vertex $v(0,0)$. Indeed, taking the inverse $Z$-transform we get:
\[
v(0,0) = \frac{1}{2\pi i}\int_{\sigma}\frac{2Ads}{\sqrt{(s-\eta_{i1})(s-\eta_{i2})}} = 2A,
\]
where $\sigma$ is a unit circle bypassed in a positive direction. Thus, $A=C_1/2$.

Similarly,
\[
\ptl_{(0,0)}[v] = \frac{A}{2\pi i}\int_{\sigma}\sqrt{(s-\eta_{o1})(s-\eta_{o2})}\frac{ds}{s} = \sqrt{\eta_{o1}\eta_{o2}} \frac{C_1}{2}.
\]
Thus, we obtain (\ref{eq:C2_C1}).

Let's now derive the embedding formula using the operator approach. Set for the sake of convenience $C_2 = 1$, and 
\[
v(0,0) = C_1 = \frac{1}{\sqrt{\eta_{o1}\eta_{o2}}}.
\]

\begin{figure}
    \centering
    \includegraphics[width=0.5\linewidth]{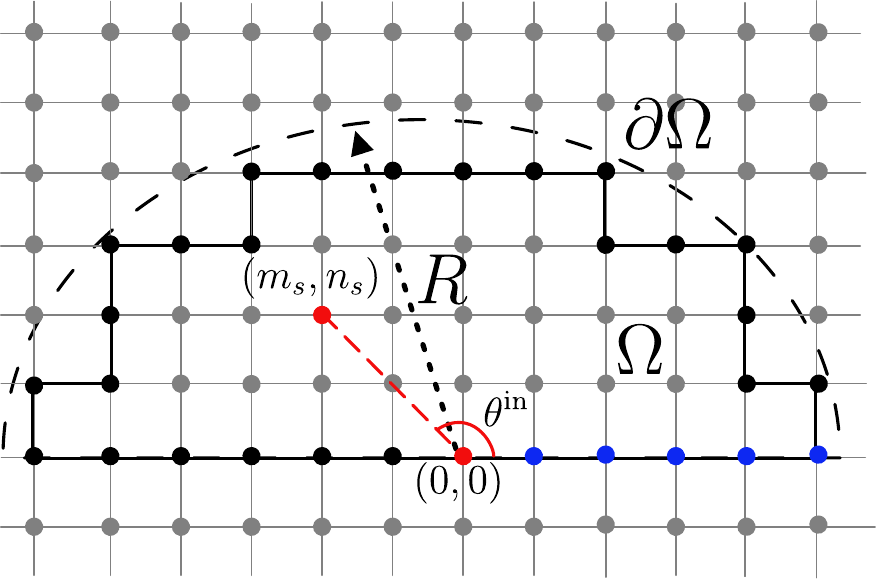}
    \caption{Domain for Green's identity for functions $v(m,n)$ and $u^{\bnu_s}$}
    \label{fig:Greens_reciprocity_edge}
\end{figure}

Study  $H^{1}_{(m,n)}[u]$. It satisfies Helmholtz equation and boundary conditions everywhere, except in the origin $(0,0)$, where 
\[
\Delta_{m,n}\left[H^{1}_{(0,0)}[u]\right] + \lk^2 H^{1}_{(0,0)}[u] = \Delta_{0,0}[u]. 
\]
Thus, 
\[
H^{1}_{(0,0)}[u] = \Delta_{0,0}[u]v(m,n).
\]
Similar to the continuos case we express $\Delta_{0,0}[u]$ through the directivity of the edge Green's function.  Indeed, consider two point source solutions of the Helmholtz equation with boundary conditions,  one is edge Green's function $v(m,n)$ and another one  is $u^{\bnu_s}$ with a source at some distant point $\bnu_s$ with amplitude $(g(m_s,n_s))^{-1}$. Applying Greens theorem to the upper half-space and taking the symmetry of the problem into account (see~Figure~\ref{fig:Greens_reciprocity_edge}) we immediately obtain:
\[
\frac{1}{2}\Delta_{0,0}[u^{\bnu_s}]v(0,0) + v(m_s,n_s)(g(m_s,n_s))^{-1} = 0.
\]
Taking the limit $\sqrt{m_s^2+n_s^2}\to \infty$ such that $m_s/n_s = \beta^{\rm in}$ we obtain:
\[
\Delta_{0,0}[u] =-\frac{2S(\beta^{\rm in})}{v(0,0)}. 
\]
Finally, strong embedding formula for directivities is:
\[
S(\beta,\beta^{\rm in}) = -2\sqrt{\eta_{o1}\eta_{o2}}\frac{S\left(\beta^{\rm in}\right)S\left(\beta\right)}{1-(s_\beta s^{\rm in})^{-1}}.
\]
Similar reasoning can be extended to the general case, however there will be no explicit expressions for the edge values of the field.

\end{document}